\pdfoutput=1
\documentclass[aps,prl,twocolumn,superscriptaddress,amsmath]{revtex4-1}
\usepackage{graphicx}
\usepackage{hyperref}
\usepackage{mathrsfs}
\usepackage{bm}
\usepackage{color}
\usepackage{siunitx}
\usepackage[capitalize]{cleveref}
\hypersetup{hypertex=true,
	colorlinks=true,
	anchorcolor=blue,
	linkcolor=blue,
    citecolor=blue,
	urlcolor=blue}
\begin{document}

\title{Universal spin superconducting diode effect from spin-orbit coupling}

\author{Yue Mao}
\affiliation{International Center for Quantum Materials, School of Physics, Peking University, Beijing 100871, China}
\author{Qing Yan}
\affiliation{International Center for Quantum Materials, School of Physics, Peking University, Beijing 100871, China}
\author{Yu-Chen Zhuang}
\affiliation{International Center for Quantum Materials, School of Physics, Peking University, Beijing 100871, China}
\author{Qing-Feng Sun}
\email[]{sunqf@pku.edu.cn}
\affiliation{International Center for Quantum Materials, School of Physics, Peking University, Beijing 100871, China}
\affiliation{Hefei National Laboratory, Hefei 230088, China}

\date{\today}

\begin{abstract}
We propose a universal spin superconducting diode effect (SDE)
induced by spin-orbit coupling (SOC), where the critical spin supercurrents in opposite directions are unequal.
By analysis from both the Ginzburg-Landau theory and energy band analysis,
we show that the spin-$\uparrow \uparrow$ and spin-$\downarrow \downarrow$ Cooper pairs possess opposite phase gradients and opposite momenta from the SOC,
which leads to the spin SDE.
Two superconductors with SOC,
a $p$-wave superconductor as a toy model and a practical superconducting nanowire,
are numerically studied and they both exhibit spin SDE.
In addition, our theory also provides a unified picture for both spin and charge SDEs.
Besides, we propose spin-polarized detection and nonlocal spin transport, as mature experimental technologies, to confirm the spin SDE in superconducting nanowires.
\end{abstract}

\maketitle

\emph{Introduction}.--Supercurrent in superconductors is an enduring research topic \cite{Josephson1962,Anderson1963,Book1964,Book1966,Book1975}.
Recent experiments have reported an exotic phenomenon that
the critical charge supercurrents in opposite directions are different,
which is called superconducting diode effect (SDE) \cite{Ono2020_SDEE,Ono2022_SDEE,Fu2022_SDEE,Li2022_SDEE,Ali2022_SDEE,Parkin2022_SDEE,Paradiso2022_SDEE}.
Such nonreciprocity has potential applications in superconducting logic circuits
and sensors with robust rectification \cite{Ono2020_SDEE,Ali2022_SDEE}.
From a theoretical perspective, the charge SDE stems from spatially modulated
order parameter $\Delta e^{iqx}$ or finite Cooper pair momentum \cite{Daido2022_SDET,Bergeret2022_SDET,Fu2022_SDET1,Fu2022_SDET2,Nagaosa2022_SDET,Hu2022_SDET,Loss2022_SDET}.

The Cooper pair, as the carrier of superconductors, has a fixed charge $2e$,
while its spin can be singlet or triplet \cite{Mackenzie2003_STSC,Birge2010_STSC,Eschrig2015_SpinSC,Blamire2018_STSC}.
Some superconductors with spin-triplet components can be managed to realize a spin supercurrent \cite{Eschrig2015_SpinSC,Grein2009_SpinSC,Alidoust2010_SpinJose,Shomali2011_SpinJose,Hikino2013_SpinJose,Linder2015_SpinJose,Eschrig2018_SpinSC,Mao2021_Spinbias,Mao2022_Spinphase,Dai2022_Spinbias,Visani2012_HM,Sanchez2022_HM,Linder2015_SP_Transport,Xiao2006_STSC},
which has received widespread attention because of its potential applications in storage devices with low power consumption \cite{Fert2008_Spin,Eschrig2015_SpinSC}.
Since charge and spin are two inseparable intrinsic degrees of Cooper pairs,
the existence of charge SDE provides the possibility of spin SDE.
Very recently, a spin SDE is theoretically suggested 
in a Fulde-Ferrell superconducting chain \cite{Zhang2023_SpinSDE}.
However, we wonder whether demanding the Fulde-Ferrell term is necessary,
and we would like to seek for a more general derivation of spin SDE.

\begin{figure}[b]
	\includegraphics[width=1\columnwidth]{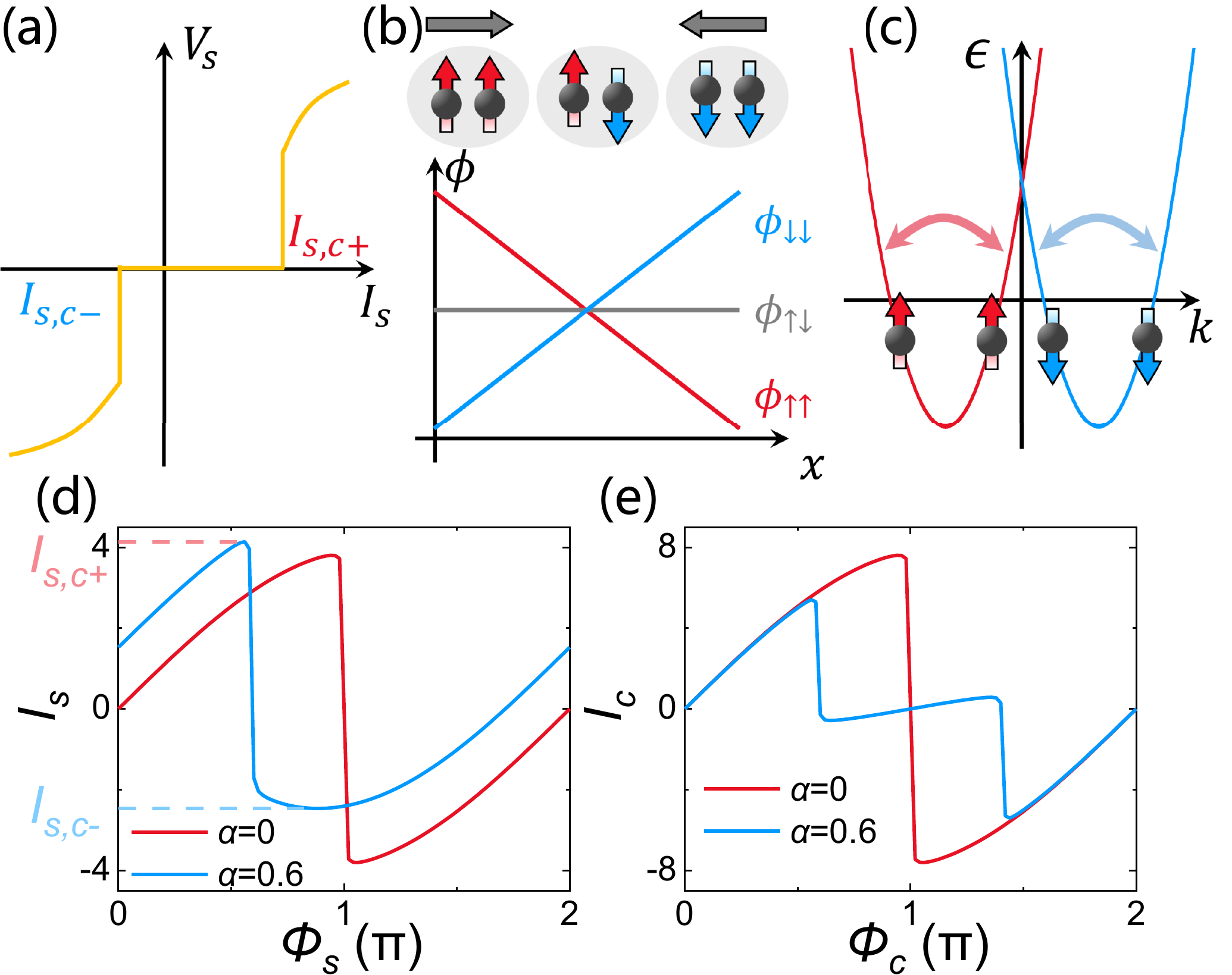}%
	\caption{\label{FIG1} (a) The schematic plot for the spin bias $V_s$
versus the spin current $I_s$ in a spin SDE device.
	(b) The SOC-induced different phase gradients
on $s_z=1,0,-1$ spin-triplet Cooper pairs (lower part),
which is equivalent to finite momenta and
causes $\uparrow\uparrow$ and $\downarrow\downarrow$ Cooper pairs
to move oppositely (upper part).
	(c) The electron energy band in the presence of SOC.
The spin $\uparrow \uparrow$ or $\downarrow \downarrow$ Cooper pairs, if existent,
will have nonzero momenta.
    (d) The spin CPR under the spin phase $\phi_s$ and
    (e) charge CPR under the charge phase $\phi_c$ in
    a $p$-wave superconductor Josephson junction.
	The critical spin supercurrents in positive and negative directions with $I_{s,c+} \ne |I_{s,c-}|$ are indicated in (d).
    Parameters: $e=\hbar=m=\Delta_p=1, \mu=10$.
	}
\end{figure}

In this Letter, we propose that the spin SDE universally appears
in the spin-triplet superconductors in the presence of spin-orbit coupling (SOC).
From both Ginzburg-Landau (GL) theory \cite{Bao2015_SSCGL1,Lv2017_SSCGL2,GL1950}
and energy band analysis, we show that the spin-$\uparrow \uparrow$ and
spin-$\downarrow \downarrow$ triplet Cooper pairs
get opposite phase gradients and momenta from SOC,
thus resulting in unequal critical spin supercurrents in positive and negative directions
[$I_{s,c+} \ne |I_{s,c-}|$, see Fig. \ref{FIG1}(a)], i.e. the spin SDE.
Then, we numerically confirm the existence of spin SDE from both
the $p$-wave superconductor as a toy model and
the practical superconducting nanowire under magnetic field.
Besides, we also give a unified physical picture for both spin SDE and charge SDE.


\emph{General theory of SOC-induced spin SDE}.--Let us
theoretically analyze the occurrence of the spin SDE
in spin-triplet superconductors while in the presence of SOC.
Without loss of generality, we below consider an one dimensional system with
the Rashba SOC $H_{\rm SO}= \alpha \sigma_z p_x $
with $\alpha$ the SOC strength \cite{Molenkamp2001_Rashba,Sun2005_Rashba}.

First, let us analyze the effect of SOC on spin transport from the GL theory:
the GL equation writes \cite{Bao2015_SSCGL1,Lv2017_SSCGL2,Mao2022_Spinphase,GL1950}
\begin{equation}
	b{\bm \Psi}+\frac{1}{2m} (-i\hbar \partial_x + 2 m \alpha \hat{s}_z )^2 {\bm \Psi} =0, \label{GL}
\end{equation}
where $ \hat{s}_z=diag(1,0,-1) $ is the $z$-direction spin operator,
$m$ is the effective mass, and
${\bm \Psi= (\Psi_{1}, \Psi_{0}, \Psi_{-1})^T}$ denotes the order parameters
of spin-triplet Cooper pair components with $s_z = 1$, $0$, and $-1$.
When there is no SOC, we can set the order parameters to be $\Psi_{s_z} (0)$.
Then one can find out that with SOC,
they become $\Psi_{s_z} (\alpha) = \Psi_{s_z} (0) e^{-2i s_z m \alpha x/\hbar}$.
We define the phases of $s_z=1, 0, -1$ Cooper pairs as $ \phi_{\uparrow \uparrow}, \phi_{\uparrow \downarrow}, \phi_{\downarrow \downarrow} $,
which are linearly dependent on the position $x$ from the above
expression of $\Psi_{s_z}(\alpha)$, as shown in Fig. \ref{FIG1}(b).
So these Cooper pairs get different phase gradients from the SOC:
$\nabla \phi_{\uparrow \uparrow}=-\frac{2m\alpha}{\hbar}, \nabla \phi_{\uparrow \downarrow}=0, \nabla \phi_{\downarrow \downarrow}=\frac{2m\alpha}{\hbar}$.
Remarkably, the phase gradients of spin-$\uparrow \uparrow$ and
spin-$\downarrow \downarrow$ Cooper pairs are opposite,
and they tend to move oppositely, see the upper part of Fig. \ref{FIG1}(b).
In fact, the phase gradient is equivalent to a Cooper pair momentum
via a unitary transformation \cite{Fu2022_SDET2}.
When the Cooper pairs as a whole get a momentum,
a charge SDE happens \cite{Fu2022_SDET2}.
Here, the opposite phase gradients of spin-$\uparrow \uparrow$
and spin-$\downarrow \downarrow$ Cooper pair components correspond to
opposite momenta $-2m\alpha$ and $2m\alpha$.
This means that $\uparrow \uparrow$ and $\downarrow \downarrow$
Cooper pairs have the same charge but
opposite spin, opposite momenta and opposite movement,
so a spin SDE should emerge.

Second, the SOC-induced spin SDE can also be obtained by
analyzing the electronic Hamiltonian with the SOC.
In the electron basis $(\psi_{k, \uparrow}, \psi_{k, \downarrow})^T$,
the one-dimensional electron system with SOC $H_{\rm SO}= \alpha \sigma_z p_x $
can be described by a simple Hamiltonian
\begin{equation}
	H_1=\frac{\hbar^2 k^2}{2m}-\mu+\alpha \sigma_z \hbar k=\frac{(\hbar k+m\alpha \sigma_z)^2}{2m}-\mu-\frac{m\alpha^2}{2},\label{H1}
\end{equation}
with the wave vector $k$.
The SOC translates spin-$\uparrow$ band by momentum $-m\alpha$
and spin-$\downarrow$ band by momentum $m\alpha$, see Fig. \ref{FIG1}(c).
If we introduce $s$-wave pairing order parameter to Eq. (\ref{H1}),
a spin-$\uparrow$ electron and a spin-$\downarrow$ electron
can combine into a spin-singlet Cooper pair with zero momentum.
However, as for spin-triplet Cooper pairs formulated by equal-spin electrons,
i.e. $\uparrow \uparrow$ and $\downarrow \downarrow$ Cooper pairs,
they will get $-2m\alpha$ and $2m\alpha$ momenta, respectively
[as shown in energy band Fig. \ref{FIG1}(c)].
It coincides with the GL analysis and a spin SDE should emerge.

\emph{Models and method}.--
In brief, above we propose that the SOC can naturally lead
to a spin SDE in spin-triplet superconductors.
Next, based on the simple SOC system in Eq. (\ref{H1}),
we add two kinds of spin-triplet superconductivity
and show the spin SDE from calculations.
One is the $p$-wave pairing as a fundamental toy model,
which gives an intelligible example for spin SDE.
The other is to apply magnetic field and proximity of $s$-wave superconductor,
which is quite mature in the field of experiments \cite{Rokhinson2012_MZME,Kouwenhoven2012_MZME,Kouwenhoven2022_MZME,Kouwenhoven2018_MZMZS}.

Next we study SDEs in the Josephson junction system where
two superconductors are connected by a short normal metal.
We would like to emphasize that the spin SDE is a bulk property
from SOC-induced Cooper pair momenta [see Sec. SI of Supplementary materials \cite{Sup}].
Here we just choose a Josephson junction to show it.
The spin and charge current-phase relations (CPRs),
i.e. $I_s-\phi_s$ and $I_c-\phi_c$, are computed using the nonequilibrium Green's function approach \cite{Sun2016_Formula,Sun2020_Formula,Sun2021_Formula},
with details provided in Supplementary Materials \cite{Sup}.
$\phi_s$ and $\phi_c$ are the spin and charge superconducting phase difference
between the left and right superconductors.
The charge phase $\phi_c$ corresponds to
a transformation $\psi'_{\uparrow,\downarrow} = \psi_{\uparrow,\downarrow} e^{i \phi_c/2}$ that leads to $\phi_{\uparrow \uparrow}=\phi_{\uparrow \downarrow}=\phi_{\downarrow \downarrow}=\phi_c$,
i.e. all Cooper pairs have the same phase and are driven in the same direction \cite{Bagwell1992_Jose,Book2001_Jose,Golubov2004_Jose}.
There are various regulations for spin Josephson supercurrent \cite{Eschrig2015_SpinSC,Alidoust2010_SpinJose,Shomali2011_SpinJose,Hikino2013_SpinJose,Linder2015_SpinJose,Mao2022_Spinphase}, which can be regarded as effectively manipulating a spin phase $\phi_s$ with $\psi_{\uparrow}' = \psi_{\uparrow} e^{i \phi_s/2}, \psi_{\downarrow}' = \psi_{\downarrow} e^{-i \phi_s/2}$ \cite{Mao2022_Spinphase}.
This spin phase corresponds to $\phi_{\uparrow \uparrow}=-\phi_{\downarrow \downarrow}=\phi_s$, $\phi_{\uparrow \downarrow}=0$ \cite{Mao2022_Spinphase,Note_Spinphase}, driving spin-$\uparrow \uparrow$ and spin-$\downarrow \downarrow$ Cooper pairs to move oppositely and induces a spin supercurrent.
After obtaining the CPRs $I_s-\phi_s$ and $I_c-\phi_c$,
the critical supercurrent in the positive (negative) direction $I_{s/c,c+}$ ($I_{s/c,c-}$) corresponds
to the maximum (minimum) current of the CPR.
The unequal critical supercurrents in opposite directions reveal the SDE \cite{Fu2022_SDET1,Fu2022_SDET2,Hu2022_SDET,Loss2022_SDET,Daido2022_SDET,Bergeret2022_SDET,Nagaosa2022_SDET}.

\emph{SDE in p-wave superconductors}.--The $p$-wave superconductivity provides
a simple toy model that the spin-triplet superconductivity and SOC coexist.
Adding this superconductivity to Eq. (\ref{H1}),
the Hamiltonian becomes $H_2=H_1+H_p$, with \cite{Balian1963_p_wave,Ueda1991_p_wave,Sarma2015_pwave}
\begin{equation}
	H_p= k (\Delta_{\uparrow \uparrow} \psi_{k,\uparrow}^\dagger \psi_{-k,\uparrow}^\dagger +\Delta_{\downarrow \downarrow} \psi_{k,\downarrow}^\dagger \psi_{-k,\downarrow}^\dagger ) + h.c.\label{Hp}
\end{equation}
Based on the above general theory between SOC and spin SDE,
as long as one of $\uparrow \uparrow$ and $\downarrow \downarrow$ components exists,
i.e. $\Delta_{\uparrow \uparrow} \ne 0 $ or $ \Delta_{\downarrow \downarrow} \ne 0$,
the spin SDE will appear.

The SOC breaks the spatial inversion symmetry, as a necessary element to realize both charge and spin SDEs.
For comparison, to realize charge SDE, the time-reversal symmetry $\mathcal{T}$ should also be broken, because the charge current is reversed by the $\mathcal{T}$ operation \cite{Hu2022_SDET}.
But a pure spin current is invariant under the $\mathcal{T}$ operation, because equivalent spin up and down components move oppositely.
Therefore, the broken $\mathcal{T}$ symmetry is unnecessary for spin SDE.
To show this, we specially calculate a $\mathcal{T}$ invariant order parameter $ \Delta_{\uparrow \uparrow}=- \Delta_{\downarrow \downarrow}=\Delta_p$.

We present the spin SDE by the spin Josephson CPR,
$I_s$-$\phi_s$,
of the $p$-wave superconductors in Fig. \ref{FIG1}(d).
When the SOC is absent, there exhibits a normal CPR with $ I_s (-\phi_s) = -I_s (\phi_s)$ and $I_{s,c+}= |I_{s,c-}|$, i.e. the spin SDE is non-existent.
The SOC $\alpha=0.6$ brings the spin SDE,
with nonreciprocal critical supercurrent $I_{s,c+} \ne |I_{s,c-}|$,
see the blue curve in Fig. \ref{FIG1}(d).
The $\phi_s$-driven spin current comes from current flow of $\uparrow \uparrow$
and $\downarrow \downarrow$ Cooper pairs $I_s (\phi_s)=\hbar[j_{\uparrow \uparrow}(\phi_s)-j_{\downarrow \downarrow} (-\phi_s)]$.
The $\mathcal{T}$ symmetry leads to the relation
$j_{\uparrow \uparrow}(\phi)=-j_{\downarrow \downarrow} (-\phi)$.
Therefore, $I_s (\phi_s)=2\hbar j_{\uparrow \uparrow}(\phi_s)=-2\hbar j_{\downarrow \downarrow} (-\phi_s)$, and $I_{s,c\pm}=2\hbar j_{\uparrow \uparrow,c\pm}=-2\hbar j_{\downarrow \downarrow,c\mp}$,
with $j_{\uparrow \uparrow (\downarrow \downarrow),c\pm}$ the critical supercurrents of $\uparrow \uparrow$ ($\downarrow \downarrow$) Cooper pairs.
The SOC-induced nonzero momenta lead to unequal critical supercurrents $j_{\uparrow \uparrow (\downarrow \downarrow),c+ } \ne |j_{\uparrow \uparrow (\downarrow \downarrow),c-}|$, and then $I_{s,c+}\ne |I_{s,c-}|$.

Different from the spin transport,
the charge SDE does not appear in Fig. \ref{FIG1}(e)
while under the drive of the charge phase $\phi_c$.
The $\phi_c$-driven charge current $I_c (\phi_c)=2e[j_{\uparrow \uparrow}(\phi_c)+j_{\downarrow \downarrow} (\phi_c)]$.
Indeed, the relation from $\mathcal{T}$ symmetry $j_{\uparrow \uparrow}(\phi)=-j_{\downarrow \downarrow} (-\phi)$ gives $I_c (-\phi_c)=-I_c (\phi_c)$, and the critical charge supercurrents in opposite directions are equal.
Therefore, the charge SDE is forbidden by the $\mathcal{T}$ symmetry.
This can also be understood by the point that
$\uparrow \uparrow$ and $\downarrow \downarrow$ Cooper pair
components each behave the SDEs,
but their contributions to charge SDE completely cancel
each other due to the $\mathcal{T}$ symmetry.

\begin{figure}
	\includegraphics[width=\columnwidth]{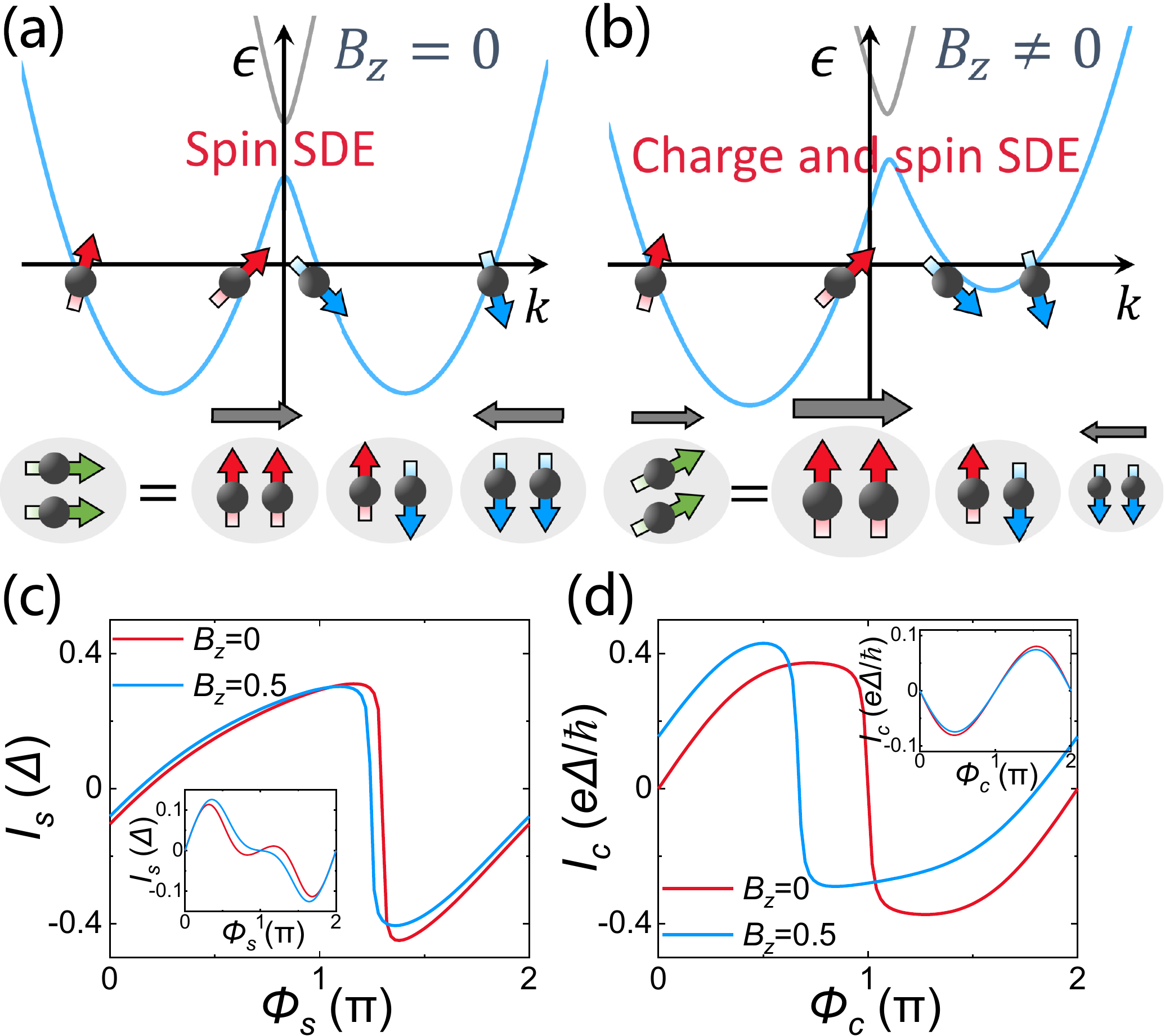}%
	\caption{\label{FIG2}
(a, b) The schematic normal-state ($\Delta=0$) energy bands of
the superconducting nanowires (upper part)
and decompositions of the dominated Cooper pairs (lower part)
with $B_x\neq 0$, and (a) $B_z=0$ and (b) $B_z \ne 0$.
(c) The CPR of $I_s$-$\phi_s$ and (d) the CPR of $I_c$-$\phi_c$
for the superconducting nanowires with the SOC strength $\alpha=0.6$.
The insets are the corresponding CPRs with $\alpha=0$.
The other parameters: $\hbar=m=\Delta=1, B_x=1.5, \mu=0$.
	}
\end{figure}

\emph{SDE of artificial spin-triplet superconductors}.--
Above we have demonstrated the spin SDE
in the $p$-wave superconductors with the SOC.
Next, we will study a more practical system, the superconducting nanowire.
Notably, mature experimental techniques \cite{Kouwenhoven2012_MZME,Rokhinson2012_MZME,Kouwenhoven2022_MZME,Kouwenhoven2018_MZMZS}
are available for fabricating superconducting nanowires
and detecting their spin-dependent transport.
The Hamiltonian of the superconducting nanowires with SOC
writes as $H_3=H_1+H_s+H_B$, where
\begin{eqnarray}
	H_s & =&\Delta \psi_{k,\uparrow}^\dagger \psi_{-k,\downarrow}^\dagger + h.c.,\label{Hs}\\
	H_B& = &(\psi_{k, \uparrow}^\dagger, \psi_{k, \downarrow}^\dagger)(-B_x \sigma_x -B_z \sigma_z)(\psi_{k, \uparrow}, \psi_{k, \downarrow})^T,\label{HB}
\end{eqnarray}
are $s$-wave superconducting pairing order Hamiltonian and magnetic field, respectively.
By combining the magnetic field and $s$-wave pairing order $\Delta$,
a dominated effective spin-triplet $p$-wave pairing is generated \cite{He2014_AR,Mao2022_Spinphase},
which will cause the spin SDE.
Without loss of generality, we set magnetic field components $B_x$ and $B_z$,
which are parallel to and perpendicular to nanowire, respectively.
For a strong magnetic field, the gap is dominated by the magnetic field,
where the spin-triplet Cooper pairs are
\emph{spin-polarized almost parallel to the magnetic field},
as shown in Figs. \ref{FIG2}(a, b) \cite{Oreg2010_MZMT,He2014_AR,DSM2015_STS,Mao2022_Spinphase}.

When the magnetic field is along direction $x$,
the dominated Cooper pairs can be decomposed into
the spin-$z$ basis $|x x\rangle= \frac{1}{2} (|\uparrow \uparrow \rangle+|\uparrow \downarrow\rangle+|\downarrow \uparrow\rangle+|\downarrow \downarrow\rangle) $.
Therefore, $\uparrow \uparrow$ and $\downarrow \downarrow$ components have
the same weight (Fig. \ref{FIG2}(a)).
This case is similar to the $p$-wave toy model.
The opposite momenta of $\uparrow \uparrow$ and $\downarrow \downarrow$ Cooper pairs
lead to spin SDE, while the charge SDE is offset.
Fig. \ref{FIG2}(c) shows the spin supercurrent $I_s$ versus
the spin phase $\phi_s$, and it clearly shows the spin SDE,
i.e. $I_{s,c+}\neq |I_{s,c-}|$, see the red curve.
But $I_c(-\phi_c) = -I_c(\phi_c)$ and there is no charge SDE,
see the red $I_c$-$\phi_c$ curve in Fig. \ref{FIG2}(d).
The disappearance of charge SDE corresponds to a symmetric normal-state energy band
in Fig. \ref{FIG2}(a), and this is consistent with some theoretical statements
that the charge SDE relates to an asymmetric band \cite{Fu2022_SDET1}.

Specially, when a $z$-direction magnetic component exists, for example $B_z > 0$,
the spin polarization is deviated from $x$ direction towards $z$ direction.
Then the $\uparrow \uparrow$ component exceeds the $\downarrow \downarrow$ component,
as shown in the lower part in Fig. \ref{FIG2}(b).
As these two components have opposite momenta,
the system still exhibits spin SDE [see the blue curve in Fig. \ref{FIG2}(c)].
Moreover, the Cooper pairs have a nonzero momentum in total,
as the charge transport is dominated by the $\uparrow \uparrow$ Cooper pairs.
As a result, the charge SDE also exists as shown by the blue curve in Fig. \ref{FIG2}(d).
Correspondingly, the normal-state energy band becomes asymmetric about $k=0$
[see the upper part in Fig. \ref{FIG2}(b)].

For comparison, in the insets of Figs. \ref{FIG2}(c, d),
the spin and charge CPRs are shown while the SOC $\alpha=0$.
There is no SDE in neither spin nor charge transport.
This implies that the SOC is a key factor causing the spin and charge SDE.

From the spin SDE perspective,
our theory provides an explanation for the charge SDE studied previously.
These charge SDEs also emerge in superconductors with both SOC and magnetic field \cite{Daido2022_SDET,Bergeret2022_SDET,Fu2022_SDET1,Loss2022_SDET,Paradiso2022_SDEE,Ono2020_SDEE},
and strongly depend on the direction of magnetic field \cite{Fu2022_SDET1,Loss2022_SDET,Paradiso2022_SDEE,Ono2020_SDEE}.
We here give an understanding that in these systems the SOC induces opposite momenta
on $\uparrow \uparrow$ and $\downarrow \downarrow$ Cooper pairs.
Their inequivalent proportions, controlled by the magnetic field, can lead to the charge SDEs.

We next investigate the efficiency of spin SDE and concentrate on the $B_z = 0$ case.
We define the spin SDE efficiency as 
\begin{equation}
	\eta=\frac{I_{s,c+}-|I_{s,c-}|}{I_{s,c+}+|I_{s,c-}|}.
\end{equation}
A nonzero $\eta$ relates to the appearance of spin SDE.

\begin{figure}
	\includegraphics[width=\columnwidth]{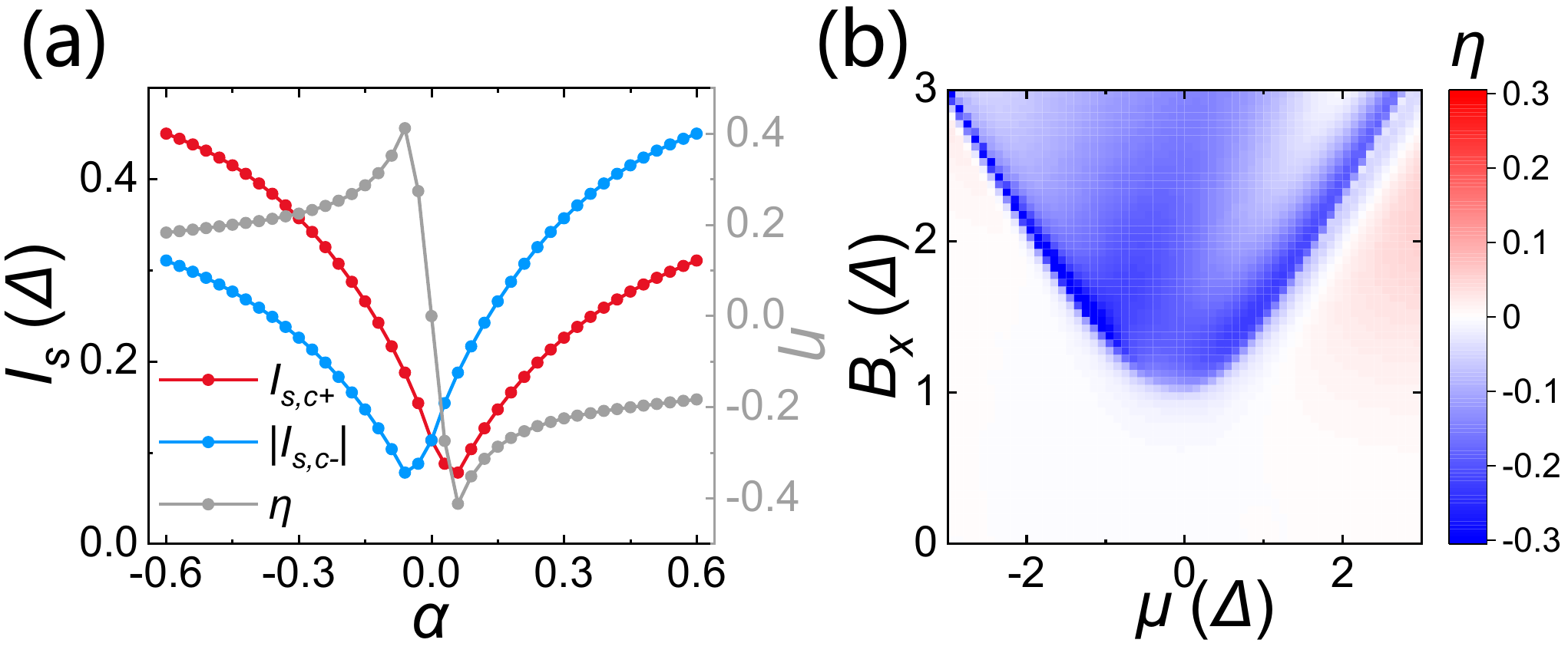}%
	\caption{\label{FIG3} 
	(a) The critical supercurrents $I_{s,c+}, |I_{s,c-}|$ and spin SDE efficiency $\eta$
as functions of SOC.
	(b) The efficiency $\eta$ as a function of magnetic field $B_x$ and chemical potential.
	Here $B_z=0$, the other parameters are the same as those in Fig. \ref{FIG2}.}
\end{figure}

We find that the spin SDE efficiency $\eta$ can be effectively regulated
by the strength of SOC $\alpha$.
Fig. \ref{FIG3}(a) shows the critical spin supercurrents in positive and negative directions versus $\alpha$.
When $\alpha<0$, 
$I_{s,c+}>|I_{s,c-}|$,
while for $\alpha>0$, $I_{s,c+}<|I_{s,c-}|$.
The Hamiltonian $H_3$ has the relation
$U H_3(\alpha) U^\dagger =H_3 (-\alpha)$ with $U=e^{-i\frac{\pi}{2}\sigma_x}$.
Meanwhile, the spin $\uparrow$ and $\downarrow$ exchange under the $U$ transformation,
thus $j_{\uparrow \uparrow}(-\alpha,\phi)=j_{\downarrow \downarrow}(\alpha,\phi)$.
Therefore, the spin CPR satisfies $I_s (-\alpha,\phi_s)= 2\hbar[ j_{\uparrow \uparrow}(-\alpha,\phi_s)-j_{\downarrow \downarrow}(-\alpha,-\phi_s) ]=2\hbar[j_{\downarrow \downarrow}(\alpha,\phi_s)-j_{\uparrow \uparrow}(\alpha,-\phi_s)]=-I_s (\alpha,-\phi_s)$,
and $I_{s,c+} (\alpha) = |I_{s,c-} (-\alpha)|$ as shown in Fig. \ref{FIG3}(a).
Correspondingly, the efficiency $\eta$ is an odd function of $\alpha$, i.e. $\eta(-\alpha)=-\eta(\alpha)$, see Fig. \ref{FIG3}(a).
The efficiency can reach $40 \% $ at $\alpha=-0.06$, which corresponds to a remarkable spin SDE with the ratio $I_{s,c+}/|I_{s,c-}| \approx 230\%$.

As the necessary factor of spin SDE, the strength of SOC is feasible.
In Fig. \ref{FIG2}, we choose a dimensionless $\alpha=0.6$.
When the superconducting gap is 250 ${\rm \mu eV}$, this SOC strength corresponds to $20\ {\rm meV \cdot nm}$, a typical value in InSb semiconductors \cite{Kouwenhoven2012_MZME,Kouwenhoven2018_MZMZS}.
Importantly, we emphasize that a large SOC is not a requirement for the spin SDE.
Even when $\alpha$ is as low as $0.06$ (equivalent to $2\ {\rm meV \cdot nm}$), the spin SDE is apparent with $ \eta \approx -40 \% $.
Therefore, the range of materials that can demonstrate spin SDE is expanded to include those with weak SOC.
Additionally, even zero-SOC systems can be regulated to exhibit spin SDE, as SOC can be added through an applied electric field \cite{Nitta1997_SOC,Koo2009_SOC,Shalom2010_SOC,Nitta2015_SOC}.

We also study the dependence of $\eta$ on magnetic field $B_x$ and chemical potential $\mu$, as shown in Fig. \ref{FIG3}(b).
As the magnetic field increases, the superconducting gap changes from a spin-singlet type (dominated by $\Delta$) to a spin-triplet type (dominated by $B_x$) \cite{Oreg2010_MZMT,Mao2022_Spinphase},
and the transition line is $B_x^2=\Delta^2+\mu^2$.
Because the spin-triplet superconductivity is the key element for spin SDE, the spin SDE is quite noticeable when the gap is $B_x$-dominated with $B_x^2>\Delta^2+\mu^2$ [see Fig. \ref{FIG3}(b)].

\emph{Detecting spin SDE in experiments}.--Based on the compelling $I_s-V_s$ curve
shown in Fig. \ref{FIG1}(a), here we suggest two practical experimental detection protocols.

\begin{figure}
	\includegraphics[width=\columnwidth]{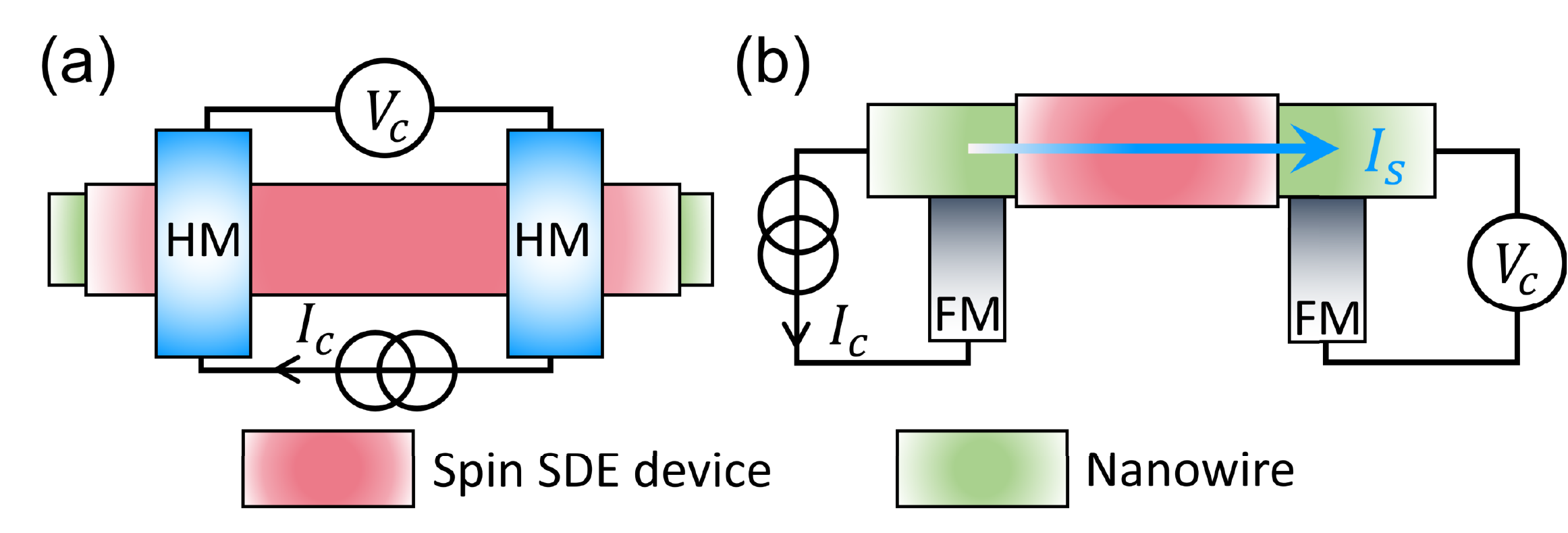}%
	\caption{\label{FIG4}
Schematic plots of the devices to detect spin SDE by
(a) spin-polarized charge transport and (b) nonlocal spin transport.
	}
\end{figure}

Half-metal ferromagnet, such as ${\rm CrO_2}$, has near $100\%$ spin polarization \cite{Korotin1998_HM,Parker2002_HM}.
Because the half-metal ferromagnet perfectly binds charge and spin together,
it can be applied to detect spin-triplet supercurrent from charge transport \cite{Xiao2006_STSC,Visani2012_HM,Sanchez2022_HM,Linder2015_SP_Transport}.
As shown in Fig. \ref{FIG4}(a), the spin SDE can be detected by local measurements
using the half-metal electrodes.
Because the spin transport is readily converted into charge signals,
the charge SDE $ I_{c,c+} \ne |I_{c,c-}| $ from $I_c-V_c$ curve
provides a faithful representation of the spin SDE $ I_{s,c+} \ne |I_{s,c-}|$
[see Fig. S1(a) in \cite{Sup}].

The other detecting method is the nonlocal spin transport.
In Fig. \ref{FIG4}(b) a spin current $I_s$ is generated
by charge injection $I_c$ from a ferromagnet (FM) electrode \cite{Tinkham2006_Ferromagnetic,Chen2013_Nonlocal,Linder2015_SP_Transport,Han2020_Spin}.
As the spin current $I_s$ flows through the spin SDE device
and arrives at the other FM electrode, it is transformed to a voltage $V_c$ \cite{Tinkham2006_Ferromagnetic,Chen2013_Nonlocal,Linder2015_SP_Transport,Han2020_Spin}.
When $I_{s,c-}<I_s<I_{s,c+}$, the spin current is a supercurrent in the spin SDE device region, and a high $\frac{dV_c}{dI_c}$ appears.
Otherwise, the spin current flows as a normal current that is dissipated in the long
distance transport, and $\frac{dV_c}{dI_c}$ sharply decreases at $I_c=I_{c,c-}, I_{c,c+}$
(i.e. at $I_s=I_{s,c-}, I_{s,c+}$) [see Fig.S1(b) in \cite{Sup}].
The signal $ I_{c,c+} \ne |I_{c,c-}| $ can also indicate the spin SDE.

\emph{Conclusion and discussion}.--In summary, we propose
that the SOC leads to opposite phase gradients
and opposite momenta on $\uparrow\uparrow$ and $\downarrow\downarrow$
spin-triplet Cooper pairs, and a universal spin SDE is caused.
For both a $p$-wave superconductor and an artificial superconducting nanowire system,
the spin Josephson CPRs with $I_{s,c+} \ne |I_{s,c-}|$
verifies the existence of spin SDE.
The spin SDE is obvious in a wide parameter space.
Our theory also provides a new perspective to view the previously studied charge SDE.

Our focus in this study is on one-dimensional systems, which provide a simplified
but insightful model for understanding the fundamental principles underlying
the spin SDE phenomenon.
However, our findings have broader implications and can be extended
to higher dimensions, allowing for a deeper understanding of the behavior
of spin-triplet superconductors.

\begin{acknowledgments}
    Y. M. is grateful to Ming Gong, Yi-Xin Dai, and Zherui Yang for fruitful discussions.
	This work was financially supported by NSF-China (Grant No. 11921005),
    the Innovation Program for Quantum Science and Technology (2021ZD0302403),
	and the Strategic priority Research Program of Chinese Academy of Sciences (Grant No. XDB28000000).
\end{acknowledgments}


\begin{references}
\bibitem{Note_Spinphase}
	Here $\phi_s$ is indeed the $z$ component of spin phase.
	For an arbitrary SOC field at $n$ direction, the spin-$nn$ and spin-$\bar{n} \bar{n}$ Cooper pairs also get opposite phase gradients, and the $n$-direction spin SDE can be found by adding an $n$-direction spin phase.
\end{references}
\end{document}


\renewcommand\thesection{S\Roman{section}}
\def\theequation{S\arabic{equation}}
\def\thefigure{S\arabic{figure}}

\title{Supplementary materials for ``Universal spin superconducting diode effect from spin-orbit coupling"}

\author{Yue Mao}
\affiliation{International Center for Quantum Materials, School of Physics, Peking University, Beijing 100871, China}
\author{Qing Yan}
\affiliation{International Center for Quantum Materials, School of Physics, Peking University, Beijing 100871, China}
\author{Yu-Chen Zhuang}
\affiliation{International Center for Quantum Materials, School of Physics, Peking University, Beijing 100871, China}
\author{Qing-Feng Sun}
\email[]{sunqf@pku.edu.cn}
\affiliation{International Center for Quantum Materials, School of Physics, Peking University, Beijing 100871, China}
\affiliation{Hefei National Laboratory, Hefei 230088, China}
\date{\today}

\maketitle
\tableofcontents

\section{How Josephson current reflects bulk superconducting diode effect}
We would like to give an explanation that the bulk superconducting diode effect (SDE) can be reflected by the current-phase relation in Josephson junctions.

In some works the (charge) SDE is confirmed in a bulk superconductor, by adding a momentum and calculating the current-momentum relations \cite{Daido2022_SDET,Bergeret2022_SDET}.
Meanwhile, in some other works the SDE can be identified by the current-phase relations in Josephson junctions \cite{Fu2022_SDET2}.
In both ways, the critical supercurrent in the positive (negative) direction corresponds to the maximum (minimum) current of the relations.

It seems that the two methods are different because in the first method the critical supercurrent is held by a certain momentum, while in the other method the critical supercurrent is held by a phase difference.
In fact, the Cooper pair momentum is equivalent to a phase gradient \cite{Fu2022_SDET2,Daido2022_SDET}.
Therefore, when two superconductors are normally connected, the Cooper pair momentum corresponds to the phase difference of the Josephson junction.
Then the SDE of Josephson junctions will be equivalent to the bulk SDE.

\section{Discretization of the Hamiltonian}
The whole Josephson junction is discretized into a tight-binding lattice with lattice constant $a=0.05$.
The Hamiltonian writes
\begin{equation}
	H_{dis}=H_L+H_R+H_N+H_C,
\end{equation}
with $H_{L(R)}$ the Hamiltonian of the left (right) superconductor, $H_N$ the Hamiltonian of the normal region, and $H_C$ the coupling term between the normal region and two superconductors.
$H_N$ and $H_C$ are always
\begin{eqnarray}
	H_N&=& \sum_{1 \le i_x \le N} \Psi_{i_x}^\dagger \hat{H}^N_0 \Psi_{i_x}
	+ \sum_{1 \le i_x \le N-1} (\Psi_{i_x}^\dagger \hat{H}^N_x \Psi_{i_x+1} + h.c.),  \nonumber \\
	H_C&=& \Psi_{0}^\dagger \hat{T} \Psi_{1} + \Psi_{N}^\dagger \hat{T} \Psi_{N+1} +h.c. .
\end{eqnarray}
Here, $\Psi_{i_x}=(\psi_{i_x, \uparrow}, \psi_{i_x, \downarrow}, \psi_{i_x, \uparrow}^\dagger, \psi_{i_x, \downarrow}^\dagger)^T$.
The onsite Hamiltonian of the normal region $\hat{H}^N_0$, the coupling inside the normal region $\hat{H}^N_x$, and the coupling between normal region and superconductors $\hat{T}$ are
\begin{equation}
	\hat{H}^N_0 = (\frac{\hbar^2}{ma^2}-\mu_N) \times diag(1, 1, -1, -1),
\end{equation}
\begin{equation}
	\hat{H}^N_x=\hat{T}=-\frac{\hbar^2}{2ma^2} \times diag(1, 1, -1, -1) \label{HT}.
\end{equation}
In the main text, we always set a short junction with a one layer normal region, i.e. $N=1$, and the chemical potential of normal region is $\mu_N=10$.

As for the superconductors, $H_L$ and $H_R$ can also be generally written as
\begin{eqnarray}
	H_{L}&=&\sum_{i_x \le 0} \Psi_{i_x}^\dagger \hat{H}^L_0 \Psi_{i_x}
	+ \sum_{i_x \le -1} (\Psi_{i_x}^\dagger \hat{H}^L_x \Psi_{i_x+1}+h.c.), \nonumber \\
	H_{R}&=& \sum_{i_x \ge N+1} \Psi_{i_x}^\dagger \hat{H}^R_0 \Psi_{i_x}
	+ \sum_{i_x \ge N+1} ( \Psi_{i_x}^\dagger \hat{H}^R_x \Psi_{i_x+1} +h.c. ).
\end{eqnarray}

For the $p$-wave superconductor case, the Hamiltonian matrices of the left superconductor are
\begin{equation}
    \hat{H}^{p, L}_0= (\frac{\hbar^2}{ma^2}-\mu) \times diag(1, 1, -1, -1),
\end{equation}
\begin{equation}
	\hat{H}^{p, L}_x=\begin{pmatrix}
		-\frac{\hbar^2}{2ma^2}-\frac{i\alpha\hbar}{2a} & 0 & -\frac{i\Delta_p}{2a} & 0 \\
		0 & -\frac{\hbar^2}{2ma^2}+\frac{i\alpha\hbar}{2a} & 0 & \frac{i\Delta_p}{2a} \\
		-\frac{i{\Delta_p}}{2a} & 0 & \frac{\hbar^2}{2ma^2}-\frac{i\alpha\hbar}{2a} & 0 \\
		0 & \frac{i{\Delta_p}}{2a} & 0 & \frac{\hbar^2}{2ma^2}+\frac{i\alpha\hbar}{2a}
	\end{pmatrix},
\end{equation}
with the $p$-wave superconducting pairing order parameter $\Delta_p$ and the SOC strength $\alpha$.

For the superconducting nanowire case, the Hamiltonians of the left superconductor are
\begin{equation}
	\hat{H}^{s, L}_0=\begin{pmatrix}
		\frac{\hbar^2}{ma^2}-\mu-B_z & -B_x & 0 & \Delta \\
		-B_x & \frac{\hbar^2}{ma^2}-\mu+B_z & -\Delta & 0 \\
		0 & -\Delta & -\frac{\hbar^2}{ma^2}+\mu+B_z & B_x \\
		\Delta & 0 & B_x & -\frac{\hbar^2}{ma^2}+\mu-B_z
	\end{pmatrix},
\end{equation}
\begin{equation}
	\hat{H}^{s, L}_x= diag(-\frac{\hbar^2}{2ma^2}-\frac{i\alpha\hbar}{2a},
	-\frac{\hbar^2}{2ma^2}+\frac{i\alpha\hbar}{2a},
	\frac{\hbar^2}{2ma^2}-\frac{i\alpha\hbar}{2a},
	\frac{\hbar^2}{2ma^2}+\frac{i\alpha\hbar}{2a}),
\end{equation}
with the chemical potential $\mu$,
the $s$-wave superconducting pairing order parameter $\Delta$,
the $x$- and $z$-direction magnetic field strength $B_x$ and $B_z$,
and the SOC strength $\alpha$.

For both cases, the Hamiltonian of the right superconductor with a charge phase $\phi_c$ is obtained by $\hat{H}^{s (p), R}_{0 (x)}= \hat{U_c} \hat{H}^{s (p), L}_{0 (x)} \hat{U_c}^\dagger$, with $\hat{U_c}= diag(e^{i\phi_c/2},e^{i\phi_c/2},e^{-i\phi_c/2},e^{-i\phi_c/2})$.
The Hamiltonian of the right superconductor with a spin phase $\phi_s$ is obtained by ${\hat{H'}^{s (p), R }_{0 (x)}}= \hat{U_s} \hat{H}^{s (p), L}_{0 (x)} \hat{U_s}^\dagger$, with $\hat{U_s}= diag(e^{i\phi_s/2},e^{-i\phi_s/2},e^{-i\phi_s/2},e^{i\phi_s/2})$.

\section{Formulas to calculate charge and spin currents}
With the tight-binding Hamiltonian, we next show the formulas to calculate the charge and spin currents, based on Refs. \cite{Sun2016_Formula,Sun2020_Formula,Sun2021_Formula}.
The charge current is
\begin{eqnarray}
	I_c & \equiv & -e \langle \frac{d(N_{L\uparrow}+N_{L\downarrow})}{dt} \rangle
	= \frac{ie}{\hbar} \langle [\sum_{i_x \le 0} ( a_{i_x \uparrow}^\dagger a_{i_x \uparrow}+a_{i_x \downarrow}^\dagger a_{i_x \downarrow} ),H] \rangle \nonumber \\
	& = & \frac{ie}{\hbar} \langle [ (a_{0\uparrow}^\dagger a_{0\uparrow}+a_{0\downarrow}^\dagger a_{0\downarrow}), \sum_{\sigma} (t a_{0\sigma}^\dagger a_{1\sigma} - t a_{0\sigma} a_{1\sigma}^\dagger)] \rangle \nonumber \\
	& = & \frac{ie}{\hbar} \sum_{\sigma} (t \langle a_{0\sigma}^\dagger a_{1\sigma} \rangle - t \langle a_{1\sigma}^\dagger a_{0\sigma} \rangle) \nonumber \\
	& = & \frac{e}{\hbar} \sum_{\sigma} [ t {\bf G}_{1e\sigma,0e\sigma}^< (t, t) - t {\bf G}_{0e\sigma,1e\sigma}^< (t, t) ] \nonumber \\
	& = & \frac{e}{\hbar} {\rm Tr} \{\Gamma_c [ {\bf G}_{10}^< (t, t) \hat{T} - \hat{T}^\dagger {\bf G}_{01}^< (t, t) ]\},
\end{eqnarray}
with $\Gamma_c= diag (1,1,-1,-1)$.
The spin current
\begin{eqnarray}
	I_s & \equiv & -\frac{\hbar}{2}\langle \frac{d(N_{L\uparrow}-N_{L\downarrow})}{dt} \rangle
	= \frac{i}{2} \langle [\sum_{i_x \le 0} ( a_{i_x \uparrow}^\dagger a_{i_x \uparrow}-a_{i_x \downarrow}^\dagger a_{i_x \downarrow} ),H] \rangle \nonumber \\
	& = & \frac{i}{2} \langle [ (a_{0\uparrow}^\dagger a_{0\uparrow}-a_{0\downarrow}^\dagger a_{0\downarrow}), \sum_{\sigma} (t a_{0\sigma}^\dagger a_{1\sigma} - t a_{0\sigma} a_{1\sigma}^\dagger)] \rangle \nonumber \\
	& = & \frac{i}{2} \sum_{\sigma} \sigma (t \langle a_{0\sigma}^\dagger a_{1\sigma} \rangle - t \langle a_{1\sigma}^\dagger a_{0\sigma} \rangle) \nonumber \\
	& = & \frac{1}{2} \sum_{\sigma} \sigma [ t {\bf G}_{1e\sigma,0e\sigma}^< (t, t) - t {\bf G}_{0e\sigma,1e\sigma}^< (t, t) ] \nonumber \\
	& = & \frac{1}{2} {\rm Tr} \{\Gamma_s [{\bf G}_{10}^< (t, t) \hat{T} - \hat{T}^\dagger {\bf G}_{01}^< (t, t) ]\},
\end{eqnarray}
with $\Gamma_s= diag (1,-1,-1,1)$.
$\hat{T}$ is the hopping matrix between the left superconductor and the normal region in Eq. (\ref{HT}).
The formulas of charge and spin currents contain the same term $ {\bf M} = {\bf G}_{10}^< (t, t) \hat{T} - \hat{T}^\dagger {\bf G}_{01}^< (t, t) $, which is to be derived next.
The lesser Green's function can be Fourier transformed to the energy space $ {\bf G}^< (t,t) = \int \frac{d \epsilon}{2\pi} {\bf G}^< (\epsilon) $, and $ {\bf M} $ becomes
\begin{eqnarray}
	 {\bf M} & = & \frac{1}{2\pi} \int d \epsilon [ {\bf G}_{10}^< (\epsilon) \hat{T} - \hat{T}^\dagger {\bf G}_{01}^< (\epsilon) ].
\end{eqnarray}
Here $ {\bf G}^< (\epsilon) = -f(\epsilon) [ {\bf G}^r (\epsilon) - {\bf G}^a (\epsilon) ] $.
The retarded Green's functions $ {\bf G}_{01}^r, {\bf G}_{10}^r $ are obtained by the Dyson equation $ {\bf G}_{01}^r ={\bf g}_{00}^r {\bf \Sigma}_{01}^r {\bf G}_{11}^r $, $ {\bf G}_{10}^r ={\bf G}_{11}^r {\bf \Sigma}_{10}^r {\bf g}_{00}^r $.
Here, ${\bf \Sigma}_{01}^r=\hat{T}, {\bf \Sigma}_{10}^r=\hat{T}^\dagger$.
${\bf g}_{00}^r$ is the surface Green's function of the left lead, which is calculated by the transformation matrix method \cite{Sun2016_Formula,Sun2021_Formula}.
${\bf G}_{11}^r$ is the Green's function of the leftmost layer of normal region ($i_x=1$), which can be obtained by a recursive algorithm \cite{Sun2016_Formula}.
The advanced Green's functions are $ {\bf G}_{10}^a =({\bf G}_{01}^r)^\dagger $, $ {\bf G}_{01}^a =({\bf G}_{10}^r)^\dagger $.
Therefore, $ {\bf M} $ can be written as
\begin{eqnarray}
	{\bf M} & = & \frac{1}{2\pi} \int d \epsilon [ -f({\bf G}_{11}^r {\bf \Sigma}_{10}^r {\bf g}_{00}^r-{\bf G}_{11}^a {\bf \Sigma}_{10}^r {\bf g}_{00}^a) {\bf \Sigma}_{01}^r
	+ f{\bf \Sigma}_{10}^r ({\bf g}_{00}^r {\bf \Sigma}_{01}^r {\bf G}_{11}^r-{\bf g}_{00}^a {\bf \Sigma}_{01}^r {\bf G}_{11}^a)] \nonumber \\
	& = & \frac{1}{2\pi} \int d \epsilon [ - f({\bf G}_{11}^r {\bf \Sigma}_{LL}^r -{\bf G}_{11}^a {\bf \Sigma}_{LL}^a)
	+ f ({\bf \Sigma}_{LL}^r {\bf G}_{11}^r-{\bf \Sigma}_{LL}^a {\bf G}_{11}^a)] \nonumber \\
	& = & \frac{1}{2\pi} \int d \epsilon ({\bf G}_{11}^r {\bf \Sigma}_{LL}^< + {\bf G}_{11}^< {\bf \Sigma}_{LL}^a -{\bf \Sigma}_{LL}^r {\bf G}_{11}^< - {\bf \Sigma}_{LL}^< {\bf G}_{11}^a )
\end{eqnarray}
Here $ {\bf \Sigma}_{LL}^r= {\bf \Sigma}_{10}^r {\bf g}_{00}^r {\bf \Sigma}_{01}^r $, $ {\bf \Sigma}_{LL}^a= {\bf \Sigma}_{10}^r {\bf g}_{00}^a {\bf \Sigma}_{01}^r $, $ {\bf \Sigma}_{LL}^<= -f({\bf \Sigma}_{LL}^r-{\bf \Sigma}_{LL}^a) $.

Finally, the charge current and spin current become

\begin{eqnarray}
	I_c & = & \frac{e}{2\pi \hbar} \int d \epsilon {\rm Tr} [ \Gamma_c ({\bf G}_{11}^r {\bf \Sigma}_{LL}^< + {\bf G}_{11}^< {\bf \Sigma}_{LL}^a -{\bf \Sigma}_{LL}^r {\bf G}_{11}^< - {\bf \Sigma}_{LL}^< {\bf G}_{11}^a ) ], \label{FinalIc}\\
	I_s & = & \frac{1}{4\pi} \int d \epsilon {\rm Tr} [ \Gamma_s ({\bf G}_{11}^r {\bf \Sigma}_{LL}^< + {\bf G}_{11}^< {\bf \Sigma}_{LL}^a -{\bf \Sigma}_{LL}^r {\bf G}_{11}^< - {\bf \Sigma}_{LL}^< {\bf G}_{11}^a ) ]. \label{FinalIs}
\end{eqnarray}

To obtain the Green's functions, an imaginary part $i \delta$ should be added \cite{Sun2016_Formula,Sun2020_Formula,Sun2021_Formula}.
In this study, we always set $\delta=0.01$.
By using Eqs. (\ref{FinalIc},\ref{FinalIs}), we numerically obtain the spin and charge SDE results in
Fig. 1(d), Fig. 1(e), Fig. 2(c), Fig. 2(d), and Fig. 3 in the main text.

In our recent work Ref. \cite{Mao2022_Spinphase}, the spin CPR of superconducting nanowire was also investigated, as an application example of the concept ``spin phase".
In fact, there already appeared the spin SDE in Ref. \cite{Mao2022_Spinphase}, but it was subtle with a low SDE efficiency $\eta$.
Here in this work, a normal region is added, of which the chemical potential $\mu_N$ can regulate the nonreciprocal spin transport and make the spin SDE more remarkable.

\section{Schematic curves of spin SDE detections}

\begin{figure}[htbp]
	\includegraphics[width=0.6\columnwidth]{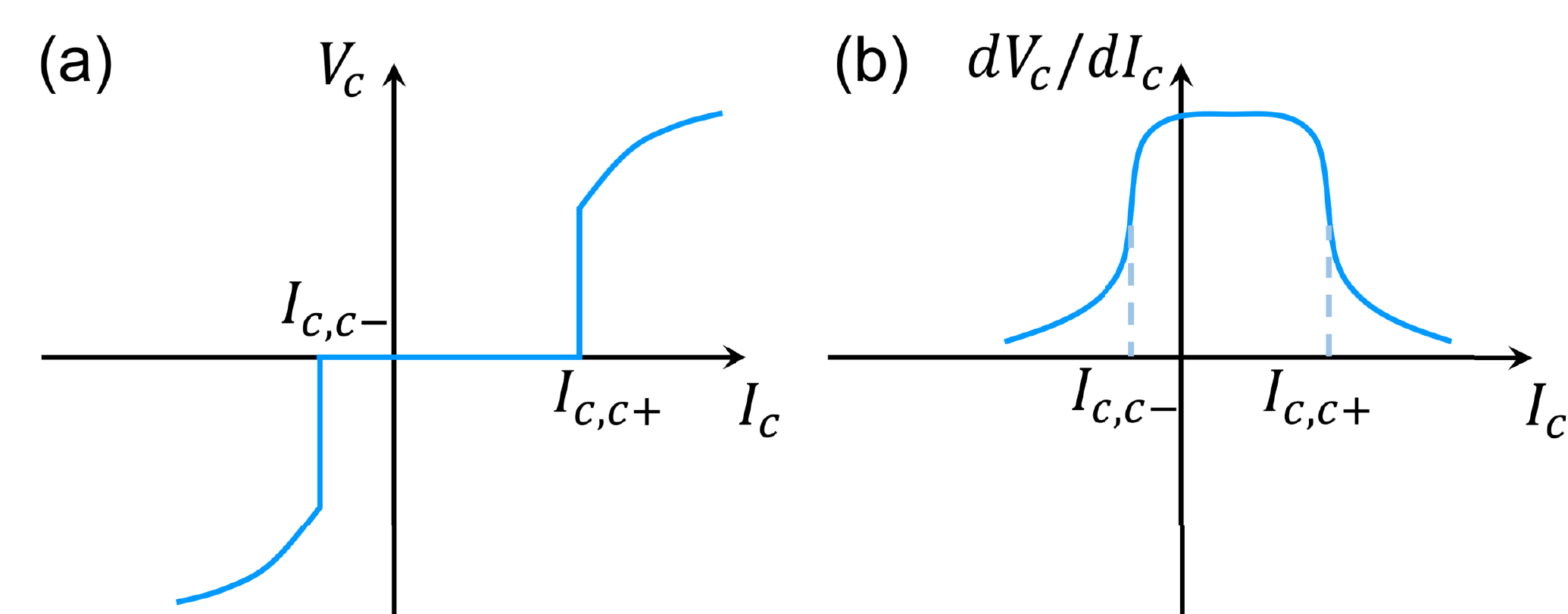}%
	\caption{\label{FIGS1} Schematic plots of charge transport signals for spin SDE in (a) spin-polarized charge transport and (b) nonlocal spin transport.
	}
\end{figure}

In Fig. \ref{FIGS1}, we show the schematic curves of charge transport to detect the spin SDE.
They correspond to spin-polarized transport and nonlocal spin transport in the main text.



